\documentclass[letter]{aa} 

\usepackage{graphicx}
\usepackage{txfonts}
\usepackage{xcolor}

\begin{document}

   \title{Extending the extinction law in 30 Doradus to the infrared  with JWST\thanks{Based on observations with the NASA/ESA {\em Hubble Space Telescope} and the NASA/ESA/CSA {\em James Webb Space Telescope}, which are operated by AURA, Inc., under NASA contracts NAS5-26555 and NAS 5-03127.}}

   \subtitle{}

   \author{Katja Fahrion\thanks{ESA research fellow}\fnmsep
          \inst{1}
          \and
          Guido De Marchi
          \inst{1}
    }

   \institute{European Space Agency (ESA), European Space Research and Technology Centre (ESTEC), Keplerlaan 1, 2201 AZ Noordwijk, the Netherlands.
             \email{katja.fahrion@esa.int, gdemarchi@esa.int}
             }

   \date{Received: February 24, 2023 / Accepted: March 8, 2023}
 
  \abstract
   {We measure the extinction law in the 30\,Dor star formation region in the Large Magellanic Cloud using Early Release Observations taken with Near-Infrared Camera (NIRCam) onboard the JWST, thereby extending previous studies with the \textit{Hubble} Space Telescope to the infrared. We use red clump stars to derive the direction of the reddening vector in twelve bands and present the extinction law in this massive star forming region from $0.3$ to $4.7\,\mu$m. At wavelengths longer than 1 $\mu$m, we find a ratio of total and selective extinction twice as high as in the diffuse Milky Way interstellar medium and a change in the relative slope from the optical to the infrared domain. Additionally, we derive an infrared extinction map and find that extinction closely follows the highly embedded regions of 30\,Dor. }


   \maketitle

\section{Introduction}
Resolved photometry of star forming regions in the Milky Way (MW) and its satellites is crucial to study stellar populations and star formation with high accuracy and to determine stellar parameters. The high spatial resolution provided by space-based telescopes such as the \textit{Hubble} Space Telescope (HST) has been vital to study the star formation process in star forming regions of our Galaxy and nearby galaxies. Photometry with HST has been used for decades to identify young stellar objects and pre-main sequence stars, and to study their spatial distributions across different star forming regions (e.g. \citealt{Gouliermis2007, DeMarchi2010, DeMarchi2011a, DeMarchi2011, Ksoll2018}). 

With a similar spatial resolution and improved sensitivity, the JWST has the potential to revolutionise our understanding of the earliest, dust-embedded phases of star formation because of its coverage in the infrared -- a wavelength regime where the extinction from interstellar dust is significantly weaker than in the optical. Indeed, the first studies using JWST observations of star forming regions have shown the potential JWST has in extending our understanding of star formation to more embedded regions and younger objects. For example, \cite{Reiter2022} used publicly available JWST images from the Early Release Observations (ERO) of the MW star formation region NGC\,3324 to identify previously undetected outflows driven by young stellar objects (YSOs), whereas \cite{Jones2023} presented JWST Near-Infrared Camera (NIRCam) observations of NGC\,346, a star formation region in the Small Magellanic Cloud, detecting low-mass, embedded YSOs with signs of accretion and dust emission.

At the same time, the high sensitivity and photometric accuracy of JWST allow us to study the extinction in the infrared towards hundreds of sight-lines, using the same techniques that have been well established for HST photometry (e.g. \citealt{DeMarchi2014a, DeMarchi2014,  DeMarchi2016_30Dor_ext, MericaJones2017, DeMarchi2021}). While the effect of dust extinction is less pronounced in the infrared, accurate measurements of stellar parameters in star forming regions will nonetheless require extinction corrections.

In this letter, we present the near-infrared (NIR) extinction law in the Tarantula nebula (30\,Doradus) in the Large Magellanic Cloud (LMC) using JWST imaging. As the closest giant HII region and the host of the candidate super star cluster NGC\,2070, 30\,Dor provides an nearby analogue to star forming regions in distant galaxies (e.g. \citealt{Crowther2019}). 30\,Dor has been the target of many photometric (e.g. \citealt{Sabbi2013, Sun2017}) and spectroscopic studies (e.g. \citealt{Evans2011, Castro2018}), covering the full wavelength range from X-ray (e.g. \citealt{Lopez2020, Crowther2022}), ultraviolet (e.g. \citealt{DeMarchi2019}), infrared (e.g. \citealt{Jones2017, Nayak2023}) to submillimeter and radio observations (e.g. \citealt{Indebetouw2013, Yamane2021, Wong2022}). Consequently, a wealth of archival data exists for 30\,Dor, including HST photometry through the Hubble Tarantula Treasury Project (HTTP, \citealt{Sabbi2013, Sabbi2016}), covering an area of 16\arcmin $\times$ 13\arcmin\, (239 pc $\times$ 194 pc assuming $D = 51.3$ kpc; \citealt{Panagia1991, Panagia2005}) with seven filters of the Wide Field Camera 3 (WFC3) and the Advanced Camera for Surveys (ACS). 
30\,Doradus was also covered with JWST as part of the ERO programme 2729 (PI: K. Pontoppidan). In this paper, we used the publicly available NIRCam imaging data of this project to obtain the extinction law in 30\,Dor and extended previous studies based on HST data (e.g. \cite{DeMarchi2016_30Dor_ext} based on the HTTP).

\section{Photometric catalogues}
\label{sect:data}
We used publicly available JWST NIRCam observations of 30\,Doradus together with archival HST data collected in the HTTP. The following briefly describes how we obtained photometric catalogues.

\begin{figure*}
    \centering
    \includegraphics[width=0.95\textwidth]{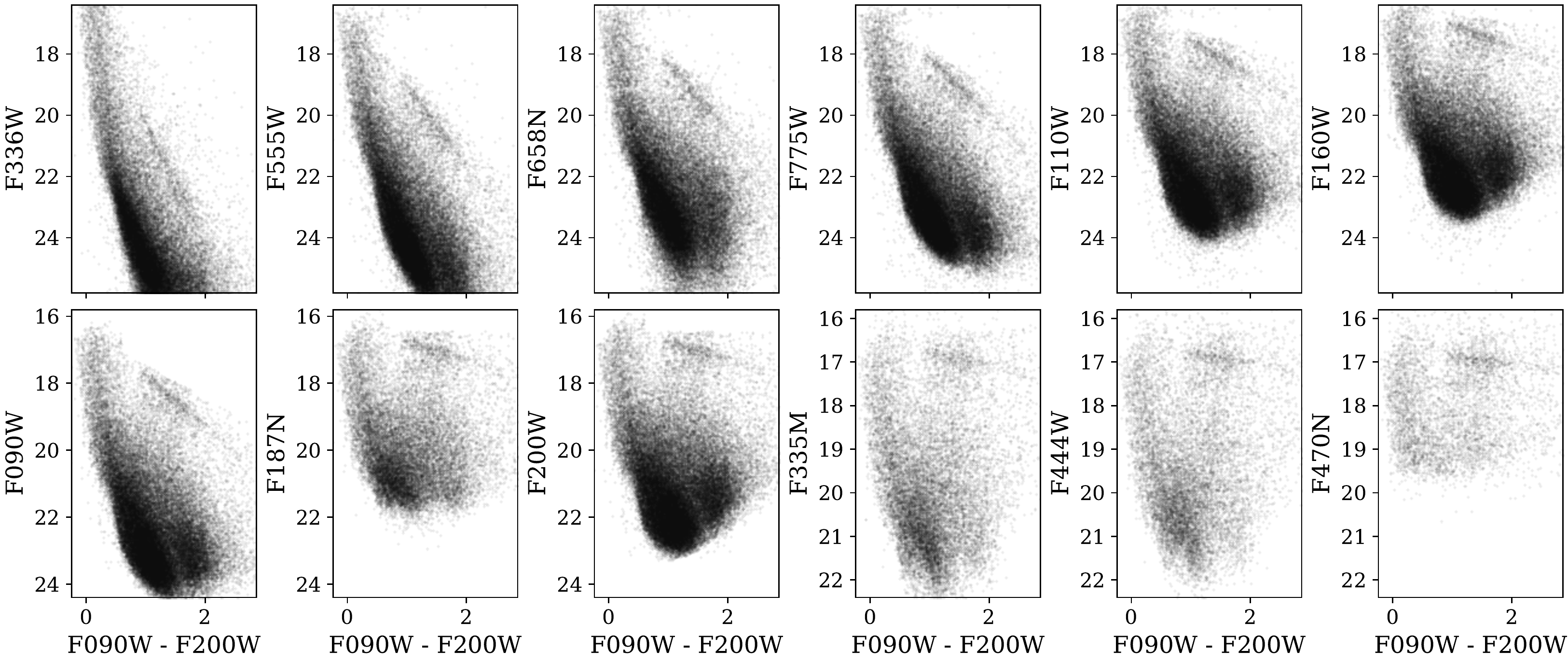}
    \caption{Colour magnitude diagrams. Each panel shows a different filter on the y-axis and F090W - F200W on the x-axis. \textit{Top row}: HST filters from the HTTP, \textit{bottom row}: NIRCam filters. The RC sequence is visible as a straight sequence brighter and redder than the main sequence.}
    \label{fig:CMDs}
\end{figure*}

\subsection{JWST data reduction}
We used observations obtained with NIRCam in the filters F090W, F187N, F200W, F335M, F444W, and F470N, spanning a 7.4\arcmin $\times$ 4.4\arcmin\,mosaic (110 pc $\times$ 66 pc), observed with a BRIGHT1 readout pattern and dithering to close the gaps between detectors. Due to the mosaic, the exposure time is not uniform across the field. In each filter, 20 individual exposures were obtained with effective exposure times of 182 seconds (F090W and F335M), 193 seconds (F200W and F444W), and 289 seconds (F187N and F470N), respectively.

To reduce the observations, we downloaded the level 1 data products from the Barbara A. Mikulski Archive for Space Telescopes (MAST)\footnote{\url{https://mast.stsci.edu}} and re-ran the stage 2 and 3 reduction steps manually. In the first step, we ran the level 2 pipeline version 1.9.3 to obtain calibrated exposures from the slope images. Then, the data were corrected for 1/f noise using the routine developed by Chris Willot\footnote{\url{https://github.com/chriswillott/jwst/}}. We used the Calibration Reference Data System mapping \textsc{jwst\_1041}.

In stage 3, we first aligned the individual exposures to Gaia data release 3 catalogues, then cosmic rays were flagged, the background was determined and finally the exposures were combined to a final mosaic. We saved the final mosaic to check the alignment between exposures and filters. The aperture photometry described in the following was done on the aligned, but uncombined level 2 outputs, so the products after the \textsc{tweakreg} step in the pipeline. We chose this intermediate pipeline stage to guarantee that all exposures are aligned to Gaia data, which ensures an accurate combination of the photometric catalogues.

\subsection{Source detection}
We used \textsc{photutil} to detect sources in the individual level 2 exposures. First, the two dimensional background of each image was subtracted using the DAOPHOT MMM algorithm. We chose a 30 by 30 pixel boxsize and a 5 by 5 pixel filter. Then, we used \textsc{photutils segmentation} routines\footnote{\url{https://photutils.readthedocs.io/en/stable/segmentation.html}} to detect sources that have at least 2 connected pixels 2$\sigma$ above the background using standard deblending parameters.

This initial source catalogue also contains saturated stars that were filtered out based on the zero flux pixel in their centre. Additionally, we applied an eccentricity cut of $\epsilon < 0.7$ to remove elongated sources such as background galaxies as well as falsely detected sources in the PSF spikes. 

\subsection{Aperture photometry}
We used \textsc{photutils aperture photometry} routines\footnote{\url{https://photutils.readthedocs.io/en/stable/aperture.html}} to perform the aperture photometry. Fluxes were obtained in apertures with 2.5 pixel radius (0.078\arcsec\, and  0.158\arcsec\, for the short and long wavelength filters, respectively) and the background in annulus apertures with inner and outer radii of 4 and 5.5 pixels, respectively, was subtracted. As the level 2 outputs are in flux units of mJy sr$^{-1}$, we used the \textsc{PIXAR\_SR} header keyword to first convert the fluxes to Jy and then to Vega magnitudes using the zeropoints as given by the SVO filter service\footnote{\url{http://svo2.cab.inta-csic.es}}. 

To convert the aperture magnitudes to total magnitudes, we used the aperture corrections provided by the JWST reference files, interpolated to an aperture radius of 2.5 pixels. 


To combine the catalogues, we only included sources that are detected in at least three exposures per filter. Here, we combined the individual measured aperture magnitudes by taking the mean and used the standard deviation as the photometric error. The final catalogue contains around 170000 sources detected in at least one NIRCam filter.

\begin{figure*}[h]
    \centering
    \includegraphics[width=0.95\textwidth]{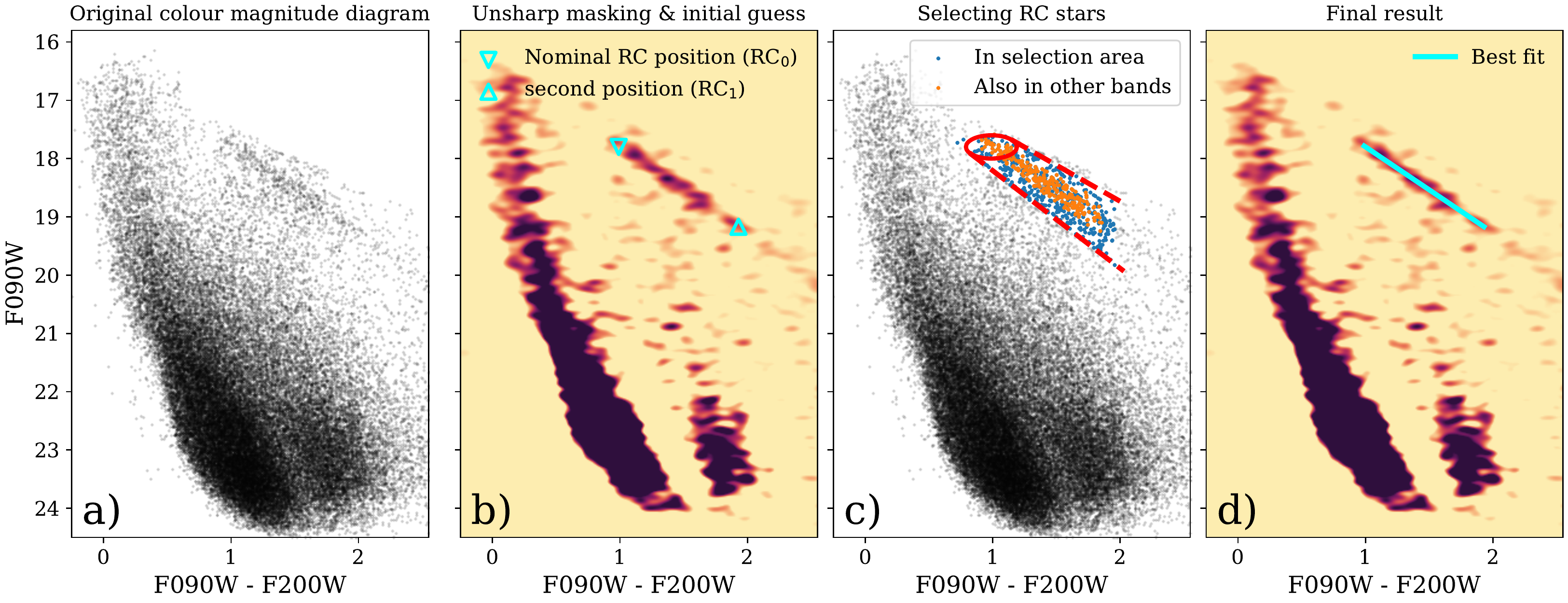}
    \caption{Illustration of RC star selection. From left to right: Panel a) shows the original F090W - F200W CMD and panel b) shows the same CMD after applying unsharp masking. We used this to obtain initial estimates for the blue and red end of the RC sequence as shown by the triangle symbols. Panel c) shows the selection of RC stars using these initial estimates to obtain a selection region. Blue dots show all stars in this CMD falling into the selection region while orange dots show the final sample of stars that fall also in similar selection regions for other filter combinations. Panel d) shows the final resulting fitted slope.}
    \label{fig:select_stars}
\end{figure*}

\subsection{Hubble Tarantula Treasury Project Data}
In addition to JWST data, we used the publicly available catalogues from the HTTP as presented by \cite{Sabbi2016}. Those include PSF photometry for stars in a 16\arcmin $\times$ 13\arcmin\, region in the HST ACS and WFC3 filters F275W, F336W, F555W, F658N, F775W, F110W, F160W. In the HTTP, both the ACS F775W and WFC3 F775W filters were used in different regions that overlap in part in the JWST NIRCam footprint. To ensure a suitable number of sources, we converted the F775W WFC3 magnitudes to F775W ACS magnitudes using stars brighter than 20.5 mag. We found a small offset of 0.012 mag between the two filters, smaller than the typical uncertainty of the RC stars. This value also matches the offset as given by PARSEC isochrones of theoretical red clump stars in these two filters \citep{Bressan2012}. 

We matched the JWST catalogue to the HST photometry from the HTTP using the filters F336W, F555W, F658N, F775W, F110W, and F160W, to connect the measured extinction back to the optical wavelength regime. As the HTTP dataset covers a much larger spatial region, we only considered the sources inside the NIRCam footprint. The HTTP also contains F275W UV data, but did not use it here due to the significantly shorter exposure time and the ``red leak'' in this filter (see \citealt{DeMarchi2016_30Dor_ext}). Figure \ref{fig:CMDs} presents colour-magnitude diagrams (CMDs) for both the HST and JWST data.

\section{Measuring extinction with red clump stars}
\label{sect:Methods}
Red clump (RC) stars are low-mass mass stars in their helium-burning phase that define a sharp feature in CMDs due their narrow range of intrinsic luminosity and colour (see the review \citealt{Girardi2016} or \citealt{Onozato2019}). They are excellent distance indicators (e.g. \citealt{Grocholski2002}), but also extinction affects their position in the CMD, making them to valuable tracers of interstellar extinction (e.g. \citealt{Nataf2013, Sanders2022}) as they scatter around the extinction vector. While the length of the so-formed RC sequence gives a handle on the maximum absolute extinction, its slope gives a direct measurement of the ratio $R$ between absolute ($A$) and selective ($E$). This slope can be influenced by the line-of-sight depth (e.g. \citealt{MericaJones2017}), but this effect should be negligible in the LMC due to the its low inclination angle ($\sim 25^{\circ}$) and small scale height ($h \sim 0.97$ kpc; \citealt{vanderMarel2001, Ripepi2022}).
In the following, we describe our approach of selecting and then fitting the slope of RC stars in the CMDs (see Figure \ref{fig:select_stars}).

\subsection{Selecting RC stars}
We inferred the nominal RC position (RC$_0$) directly from the CMD as the blue, bright end of the elongated sequence. This is the position where unextinguished RC stars should be located.
We used the unsharp masking technique to visually identify the RC, a similar approach as described in \cite{DeMarchi2016_30Dor_ext, DeMarchi2021}. Here, the CMD is first binned in a two-dimensional array with a small binsize of 0.01 mag. Then, this histogram is convolved with a narrow, two-dimensional Gaussian to assign each point a resolution that corresponds to the typical photometric uncertainty. We used 0.05 mag for the standard deviation of this Gaussian. In a second step, a broader Gaussian with a standard deviation of 0.2 mag is used to convolve this smoothed histogram even further and then the the residual image is obtained by subtracting both blurred images from another. With this technique, features such as the RC become visible.

We used these images to get an initial estimate of the slope $m_\text{init}$ by identifying RC$_0$ and a second, redder point (RC$_1$) where the density of RC stars drops, however the RC sequence can extend to redder colours (see Fig. \ref{fig:select_stars}). To then select RC stars, we adopted the approach detailed in \cite{DeMarchi2014}: we selected all stars that are either in a error ellipse around the selected nominal RC position using conservative magnitude and colour uncertainties of 0.2 mag, respectively, or inside a region bounded by the tangents to this error ellipse with slopes $m_\text{init}$ $\pm$ 15\%, extending past RC$_1$. Allowing for 15\% uncertainty to the initial slope ensures that the selected sample is not biased with respect to the initial selection.  

Using this selection yields an initial sample of RC stars for each filter (panel c in Fig. \ref{fig:select_stars}). To make the selection more robust against outliers, we considered only stars that were in the RC sample of at least ten of the twelve bands. At least 150 RC stars per filter are in this selection, with typical uncertainties of 0.02 - 0.04 mag.

\begin{table}[]
    \centering
    \caption{Measured ratios of absolute and selective extinction.}
    \begin{tabular}{l c c c c} \hline \hline
    Filter & $\lambda_{\text{eff}}$ & $\lambda_{\text{ref}}^{-1}$ & $R_\text{F090W - F200W}$ & $R_\text{F555W - F775W}$ \\
     & ($\mu$m) & ($\mu$m$^{-1}$) & & \\ \hline
F336W  & 0.340 & 2.944 & 3.72 $\pm$ 0.28 & 4.69 $\pm$ 0.25\\
F555W  & 0.542 & 1.846 & 2.58 $\pm$ 0.14 & 3.29 $\pm$ 0.18\\
F658N  & 0.658 & 1.519 & 2.17 $\pm$ 0.13 & 2.73 $\pm$ 0.19\\
F775W  & 0.770 & 1.298 & 1.79 $\pm$ 0.12 & 2.31 $\pm$ 0.18\\
F110W  & 1.152 & 0.868 & 1.08 $\pm$ 0.11 & 1.38 $\pm$ 0.15\\
F160W  & 1.536 & 0.651 & 0.70 $\pm$ 0.11 & 0.90 $\pm$ 0.13\\ \hline
F090W  & 0.900 & 1.111 & 1.47 $\pm$ 0.11 & 1.87 $\pm$ 0.17\\
F200W  & 1.968 & 0.508 & 0.49 $\pm$ 0.11 & 0.63 $\pm$ 0.12\\
F187N  & 1.874 & 0.534 & 0.54 $\pm$ 0.12 & 0.69 $\pm$ 0.13\\
F335M  & 3.350 & 0.299 & 0.26 $\pm$ 0.11 & 0.35 $\pm$ 0.11\\
F444W  & 4.364 & 0.229 & 0.22 $\pm$ 0.11 & 0.29 $\pm$ 0.11\\
F470N  & 4.708 & 0.212 & 0.20 $\pm$ 0.10 & 0.28 $\pm$ 0.11\\
\hline
    \end{tabular}
         \tablefoot{The quoted uncertainties refer to the 1$\sigma$ scatter around the residual of the best-fit. Formal errors are five times smaller.}
    \label{tab:results}
\end{table}

\subsection{Obtaining the extinction}
We then fitted the slope of the RC stars with a straight line using a least squares approach. To ensure the fits are stable against selection effects and photometric uncertainty, we fitted each sample 10000 times with a Monte Carlo approach. In each fit, the colour and magnitude of all RC stars are sampled randomly assuming Gaussian distributions for these parameters with standard deviations given by the photometric uncertainty. Additionally, we removed randomly 50\% of the stars in each fit. We then defined the best-fitting slope as the mean of the resulting parameter distribution. As uncertainty, we used the scatter of the residual. 

Formally, the scatter around the best-fit value from the Monte Carlo approach might be used, which is typically around $\sim$ 0.02. However, as discussed in \cite{DeMarchi2016_30Dor_ext}, the dispersion around the best fit is a systematic effect due to varying grain sizes and a variable extinction law across the area. Additional scatter could be introduced by variations in the line-of-sight distance and stellar populations \citep{Onozato2019}.

\begin{figure}
    \centering
    \includegraphics[width=0.48\textwidth]{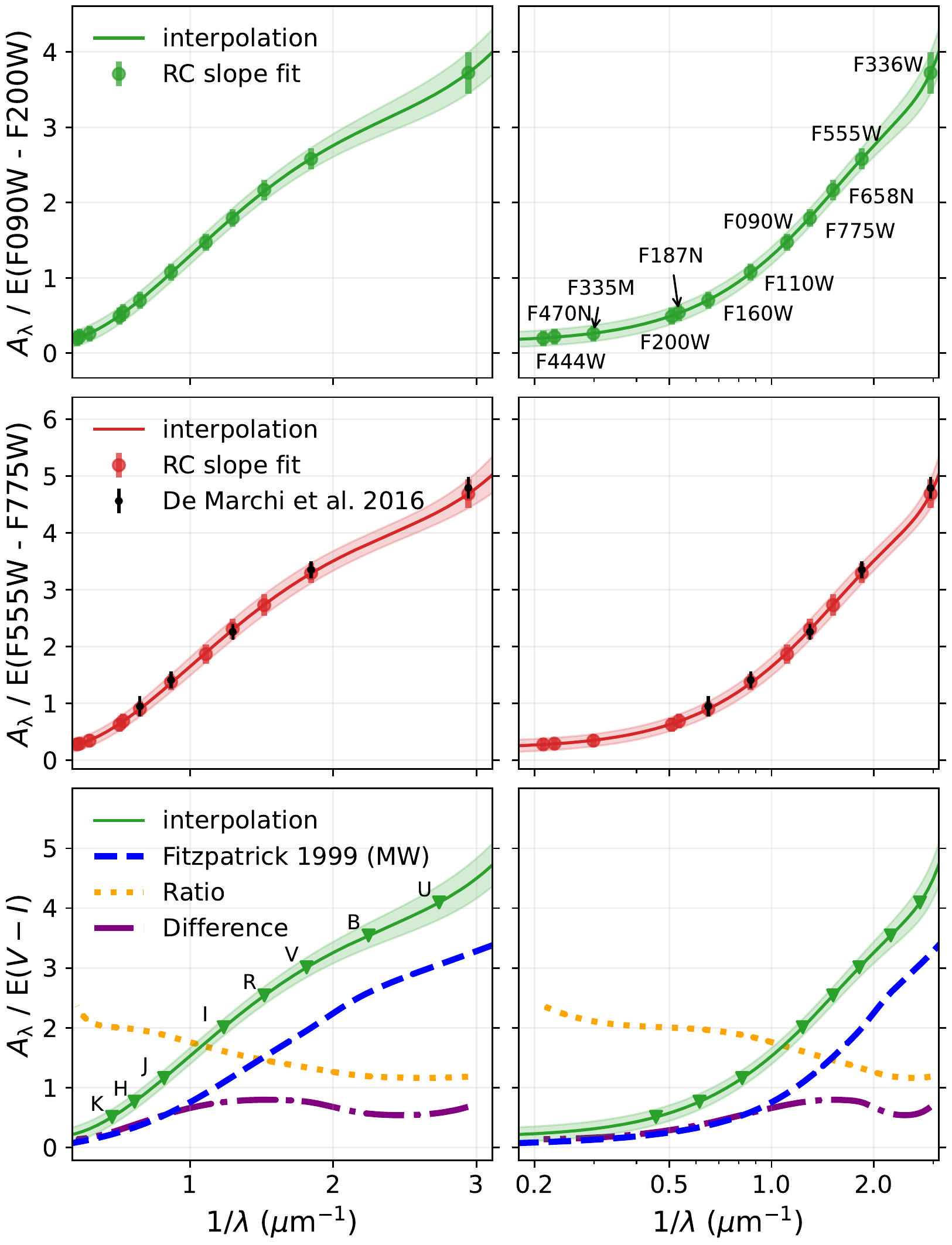}
    \caption{Extinction law with linear (\textit{left)} and logarithmic (\textit{right}) x-axis. \textit{Top}: Green dots show the measured slopes of the RC stars using the F090W - F200W JWST colour. The solid line shows a univariate spline interpolation. \textit{Middle}: Red dots show the extinction curve as a function of F555W - F775W colours with results from \cite{DeMarchi2016_30Dor_ext} (black dots) for optical wavelengths. \textit{Bottom}: Interpolated results for $A_\lambda$/E($V - I$), compared to the MW extinction law \citep{Fitzpatrick1999} with $R_V = 3.1$, corresponding to $R_{V - I}(V) = 2.3$. Ratio and difference between those curves are shown by the orange dotted and purple dash-dotted lines, respectively.}
    \label{fig:ext_law}
\end{figure}

\section{Results}
\label{sect:results}
We fitted the RC slopes in CMDs using different filter combinations. Those give the ratio $R$ between absolute ($A$) and selective ($E$) extinction at different wavelengths $\lambda$:
\begin{equation}
    R_\mathrm{1 - 2}(\lambda) = \frac{A_\lambda}{E\mathrm{(Filter_1} - \mathrm{Filter_2)}}.
\end{equation}
We used both the F090W - F200W JWST and the F555W - F775W HST colours.

\subsection{Extinction law}
Figure \ref{fig:ext_law} shows the extinction curve using the F090W - F200W and F555W - F775W colours. In the latter, we compare to the measurements presented in \cite{DeMarchi2016_30Dor_ext} using the same HTTP HST data, but on a much larger area. As can be seen, our measurements fully reproduce the literature results, but now also include the much redder JWST filters and the narrow band filters F658N, F187N, and F470N. Table \ref{tab:results} reports the best-fitting extinction ratios for each filter. The extinction curves when using F090W - F200W or F555W - F775W colours fully agree.

The solid lines in Fig. \ref{fig:ext_law} show univariate spline interpolation of degree four through the measurements and shaded areas refer to interpolations through the measurements plus or minus uncertainties. Those interpolations were used to infer extinction values for other JWST NIRCam filters (see Table \ref{tab:NIRCam_values}). 

The bottom panel of Fig. \ref{fig:ext_law} shows the extinction law as $A_\lambda$/E($V - I$) (see also \citealt{DeMarchi2016_30Dor_ext}), compared with the MW curve from \cite{Fitzpatrick1999}. Based on extrapolations beyond the $H$ band, \cite{DeMarchi2016_30Dor_ext} suggested that at wavelength longer $> 1\,\mu$m the extinction law in 30\,Dor tapers off twice as quickly as the Galactic law because of the larger proportion of large grains in the 30\,Dor interstellar medium adding a grey component to the extinction in the optical (and in the ultraviolet, \citealt{DeMarchi2019}). With JWST, we can now confirm and extend this,  finding the ratio to be relatively constant between 1 and 5 $\mu$m, suggesting the grain composition to be similar to that in the MW, but with a higher fraction of large grains. The ratio decreases for shorter wavelengths and reaches again constant values at $\lambda \sim\,0.5\,\mu$m. For a discussion on grain sizes, we refer to \cite{DeMarchi2019} that presented the extension of the 30\,Dor extinction law into the ultraviolet.

\begin{figure}
    \centering
    \includegraphics[width=0.48\textwidth]{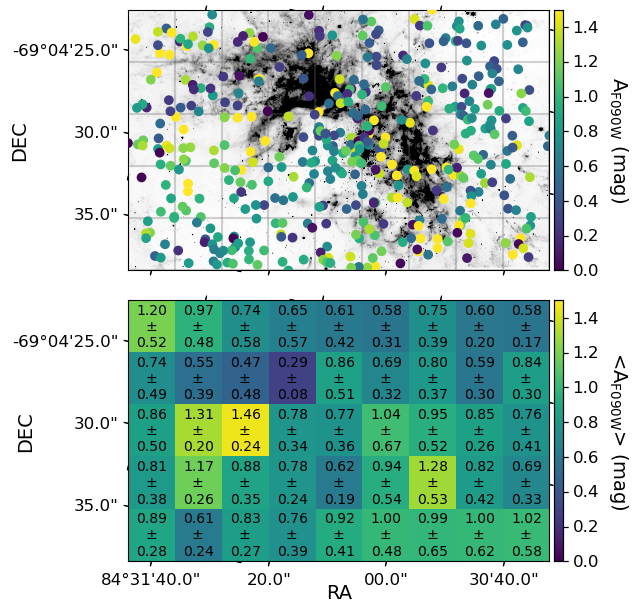}
    \caption{Extinction map. \textit{Top}: F335M mosaic of 30\,Doradus (greyscale image) with individual RC stars (dots, colour-coded by their F090W extinction). \textit{Bottom}: Average F090W extinction in a grid of the same field-of-view. The mean extinction and the standard deviation are given by the number in each cell. The number of stars in each cell ranges from five to 28.}
    \label{fig:ext_map}
\end{figure}
\subsection{Extinction map}
\label{sect:extinction_map}

Figure \ref{fig:ext_map} shows the spatial distribution of RC stars on top of the F335M image, coloured by their absolute F090W extinction. The bottom panel presents an average extinction map and notes the average extinction and the standard deviation in each of the cells. For this figure, we redid the selection of RC stars using the derived slopes, a smaller uncertainty of 0.05 mag for the error ellipse and no limitation at the red end.

The stars with highest extinction are located in and next to the web-like emission in the F335M image that follows the cold molecular gas (e.g. \citealt{Wong2022}).
However, the cell with the lowest extinction in the field is located directly on top of a F335M-bright region. We attribute this is due to low numbers and selection effects we only used RC stars detected in  F090W. In this highly embedded region, likely only foreground stars could be detected in this filter. 


\section{Discussion and conclusions}
\label{sec:discussion}

\begin{figure}
    \centering
    \includegraphics[width=0.48\textwidth]{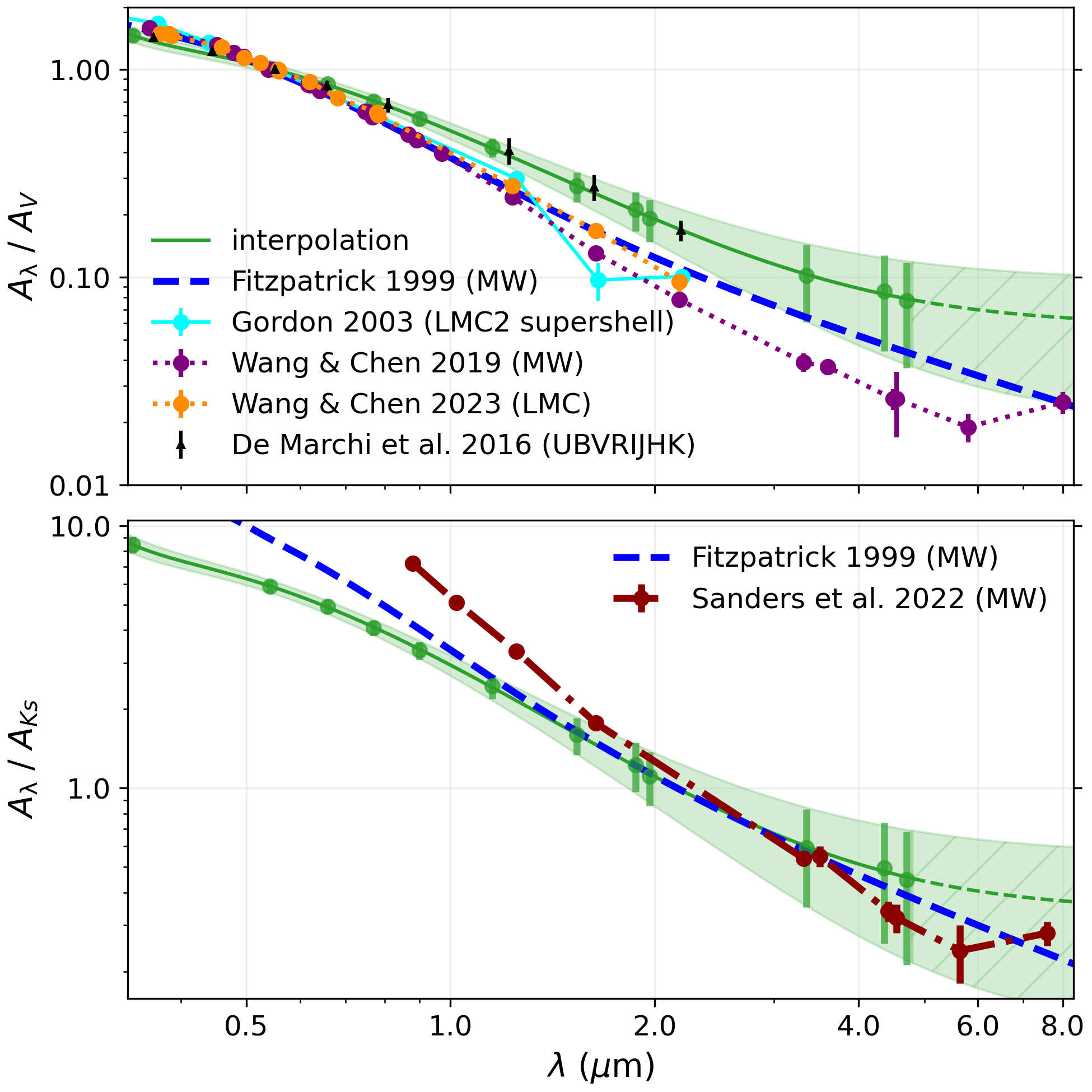}
    \caption{Comparison of relative extinction with literature results. \textit{Top}: Extinction relative to $A_V$ assuming $R_\text{F090W - F200W}$($A_V$) = 2.53 in green. Results for the MW in blue dashed \citep{Fitzpatrick1999} and purple \citep{WangChen2019}. Results for the LMC in cyan \citep{Gordon2003} and orange \citep{WangChen2023}. Black show the UBVRIHJK values from \cite{DeMarchi2016_30Dor_ext}. \textit{Bottom}: Relative to $A_{Ks}$ assuming $R_\text{F090W - F200W}$($A_V$) = 0.45. Results from \cite{Sanders2022} for the MW in dark red, and from \cite{Fitzpatrick1999} in blue.}
    \label{fig:literature_results}
\end{figure}

Figure \ref{fig:literature_results} shows relative extinction curves expressed as $A_\lambda/A_V$ and $A_\lambda/A_{Ks}$ using interpolated values for the $V$ (0.551 $\mu$m) and 2MASS $Ks$ (2.156 $\mu$m) bands. The top panel compares to literature results from the MW \citep{Fitzpatrick1999, WangChen2019}, and the LMC \citep{Gordon2003, WangChen2023}.
The bottom panel compares to the central 3 $\times$ 3 deg$^2$ of the MW from \cite{Sanders2022}.

As also noted by \cite{WangChen2023}, the extinction curve presented here and in \cite{DeMarchi2016_30Dor_ext} is generally flatter than what is found for the average curves of much wider areas. This is likely caused by spatial variations in the grain size distribution and their relative abundances. As an exemplary massive and energetic star formation region, 30\,Doradus is known to have an anomalous extinction law, as also discussed in \cite{DeMarchi2014, MaizApellaniz2014, DeMarchi2016_30Dor_ext}, or \cite{DeMarchi2019}. 
Here, we could show that this elevated extinction amplitude extends into the infrared and could confirm a transition in the slope at 1 $\mu$m.

For wavelengths $>$ 1 $\mu$m, the slope in relative $A_\lambda/A_{Ks}$ appears to follow the Galactic extinction law, implying a similar distribution of grain sizes. Our results might suggest a flattening of the curve $> 4\,\mu$m, but more measurements in the mid-infrared are needed to confirm this. The comparison of relative extinction curves reiterates what the comparison in Fig. \ref{fig:ext_law} shows: the slope of the curve derived here matches the Galactic extinction laws at long wavelengths, but flattens out at shorter wavelengths. This highlights that while the nature of the grains is likely similar as in the MW, the size distribution differs with 30\,Doradus having a larger fraction of large grains.

To conclude, this paper presents an extension of the extinction law in 30\,Doradus to the infrared, exploiting the high resolving power of JWST. By combining the JWST NIRCam data with archival HST data, we derive a uniform extinction law from $0.3$ to $4.7\,\mu$m using a fully consistent method based on the same field and the same stars. Compared with established extinction laws of the MW, we found both a higher value of $R_{V-I}$, and a change of the relative slope at $\sim$ 1 $\mu$m. At longer wavelengths, the ratio between curves is constant at $\sim$ 2, but reaches towards $\sim$ 1.2 for wavelengths shorter than $\sim$ 0.5 $\mu$m, consistent with the presence of larger grains. 
In addition to the extinction law, we presented the spatial distribution of RC stars in the JWST NIRCam field-of-view and their absolute extinction in the F090W filter, finding that the most extinguished stars are in the highly embedded regions.

With its high star formation activity and low metallicity, 30\,Doradus is often seen as a local analogue of high redshift star forming clumps. Consequently, the anomalous extinction law presented here has implications when interpreting results from high redshift, in particular as intrinsic fluxes might be underestimated when the canonical MW laws are used.

\begin{acknowledgements}
      We thank the anonymous referee for useful comments that helped polish this work.
      KF thanks Sam Pearson and Victor See for helpful discussions.
      KF acknowledges support through the ESA research fellowship programme. This work made use of Astropy:\footnote{\url{http://www.astropy.org}} a community-developed core Python package and an ecosystem of tools and resources for astronomy \citep{astropy2013, astropy2018, astropy2022}
\end{acknowledgements}

\bibliographystyle{aa} 
\bibliography{References}

\appendix
\section{Interpolated values for other JWST NIRCam filters}
Table \ref{tab:NIRCam_values} reports the ratio of absolute and selective extinction for other NIRCam filters. These values are based on the univariate spline interpolation (see Sect. \ref{sect:results}). The uncertainties are based on the envelope curves using interpolation of the measured values plus or minus the measured dispersion around the residual. For interpolated values in common optical filters, we refer to \cite{DeMarchi2016_30Dor_ext}.

\begin{table*}[b]
    \centering
    \caption{Interpolated extinction values for different JWST NIRCam filters.}
    \begin{tabular}{l c c c c c c c} \hline\hline
    Filter & $\lambda_{\text{eff}}$ & $\lambda_{\text{eff}}^{-1}$ & $R_\text{F090W - F200W}$ & $R_\text{F555W - F775W}$ & $R_{V - I}$ & $A_\lambda/A_V$ & $A_\lambda/A_{Ks}$ \\
     & ($\mu$m) & ($\mu$m$^{-1}$) & & & &\\ \hline
F070W  &  0.697 &  1.435 &  2.023 $\pm$  0.123 &  2.575 $\pm$  0.182 &  2.388 $\pm$  0.145 &  0.794 $\pm$  0.048 &  4.604 $\pm$  0.281\\
F090W  &  0.900 &  1.111 &  1.486 $\pm$  0.117 &  1.893 $\pm$  0.169 &  1.754 $\pm$  0.138 &  0.583 $\pm$  0.046 &  3.382 $\pm$  0.266\\
F115W  &  1.149 &  0.871 &  1.068 $\pm$  0.115 &  1.361 $\pm$  0.152 &  1.260 $\pm$  0.136 &  0.419 $\pm$  0.045 &  2.430 $\pm$  0.262\\
F140M  &  1.404 &  0.712 &  0.802 $\pm$  0.114 &  1.025 $\pm$  0.138 &  0.946 $\pm$  0.134 &  0.315 $\pm$  0.045 &  1.824 $\pm$  0.260\\
F150W  &  1.497 &  0.668 &  0.731 $\pm$  0.114 &  0.936 $\pm$  0.134 &  0.863 $\pm$  0.134 &  0.287 $\pm$  0.045 &  1.664 $\pm$  0.259\\
F162M  &  1.626 &  0.615 &  0.650 $\pm$  0.113 &  0.834 $\pm$  0.129 &  0.767 $\pm$  0.134 &  0.255 $\pm$  0.044 &  1.479 $\pm$  0.258\\
F164N  &  1.644 &  0.608 &  0.639 $\pm$  0.113 &  0.820 $\pm$  0.129 &  0.755 $\pm$  0.133 &  0.251 $\pm$  0.044 &  1.455 $\pm$  0.258\\
F150W2  &  1.556 &  0.643 &  0.692 $\pm$  0.113 &  0.886 $\pm$  0.132 &  0.817 $\pm$  0.134 &  0.272 $\pm$  0.044 &  1.574 $\pm$  0.258\\
F182M  &  1.839 &  0.544 &  0.546 $\pm$  0.112 &  0.703 $\pm$  0.124 &  0.644 $\pm$  0.133 &  0.214 $\pm$  0.044 &  1.242 $\pm$  0.256\\
F187N  &  1.874 &  0.534 &  0.532 $\pm$  0.112 &  0.685 $\pm$  0.123 &  0.628 $\pm$  0.132 &  0.209 $\pm$  0.044 &  1.210 $\pm$  0.256\\
F200W  &  1.968 &  0.508 &  0.497 $\pm$  0.112 &  0.642 $\pm$  0.121 &  0.587 $\pm$  0.132 &  0.195 $\pm$  0.044 &  1.131 $\pm$  0.255\\
F210M  &  2.092 &  0.478 &  0.457 $\pm$  0.111 &  0.592 $\pm$  0.119 &  0.540 $\pm$  0.131 &  0.180 $\pm$  0.044 &  1.041 $\pm$  0.254\\
F212N  &  2.121 &  0.471 &  0.449 $\pm$  0.111 &  0.582 $\pm$  0.118 &  0.530 $\pm$  0.131 &  0.176 $\pm$  0.044 &  1.022 $\pm$  0.254\\
F250M  &  2.501 &  0.400 &  0.363 $\pm$  0.110 &  0.475 $\pm$  0.114 &  0.429 $\pm$  0.130 &  0.143 $\pm$  0.043 &  0.827 $\pm$  0.250\\
F277W  &  2.741 &  0.365 &  0.326 $\pm$  0.109 &  0.429 $\pm$  0.112 &  0.385 $\pm$  0.129 &  0.128 $\pm$  0.043 &  0.742 $\pm$  0.249\\
F300M  &  2.986 &  0.335 &  0.296 $\pm$  0.108 &  0.392 $\pm$  0.111 &  0.350 $\pm$  0.128 &  0.116 $\pm$  0.042 &  0.674 $\pm$  0.247\\
F323N  &  3.237 &  0.309 &  0.272 $\pm$  0.107 &  0.363 $\pm$  0.110 &  0.321 $\pm$  0.127 &  0.107 $\pm$  0.042 &  0.620 $\pm$  0.245\\
F322W2  &  3.051 &  0.328 &  0.289 $\pm$  0.108 &  0.384 $\pm$  0.110 &  0.342 $\pm$  0.128 &  0.114 $\pm$  0.042 &  0.659 $\pm$  0.246\\
F335M  &  3.350 &  0.299 &  0.263 $\pm$  0.107 &  0.352 $\pm$  0.109 &  0.311 $\pm$  0.126 &  0.103 $\pm$  0.042 &  0.599 $\pm$  0.244\\
F356W  &  3.509 &  0.285 &  0.252 $\pm$  0.107 &  0.338 $\pm$  0.109 &  0.297 $\pm$  0.126 &  0.099 $\pm$  0.042 &  0.573 $\pm$  0.243\\
F360M  &  3.611 &  0.277 &  0.245 $\pm$  0.106 &  0.330 $\pm$  0.109 &  0.290 $\pm$  0.126 &  0.096 $\pm$  0.042 &  0.559 $\pm$  0.243\\
F405N  &  4.053 &  0.247 &  0.223 $\pm$  0.105 &  0.303 $\pm$  0.108 &  0.263 $\pm$  0.124 &  0.088 $\pm$  0.041 &  0.507 $\pm$  0.240\\
F410M  &  4.072 &  0.246 &  0.222 $\pm$  0.105 &  0.302 $\pm$  0.108 &  0.262 $\pm$  0.124 &  0.087 $\pm$  0.041 &  0.506 $\pm$  0.240\\
F430M  &  4.278 &  0.234 &  0.214 $\pm$  0.105 &  0.293 $\pm$  0.108 &  0.253 $\pm$  0.124 &  0.084 $\pm$  0.041 &  0.487 $\pm$  0.239\\
F444W  &  4.364 &  0.229 &  0.211 $\pm$  0.105 &  0.289 $\pm$  0.108 &  0.249 $\pm$  0.124 &  0.083 $\pm$  0.041 &  0.481 $\pm$  0.239\\
F460M  &  4.631 &  0.216 &  0.203 $\pm$  0.104 &  0.280 $\pm$  0.108 &  0.240 $\pm$  0.123 &  0.080 $\pm$  0.041 &  0.462 $\pm$  0.238\\
F466N  &  4.655 &  0.215 &  0.202 $\pm$  0.104 &  0.279 $\pm$  0.108 &  0.239 $\pm$  0.123 &  0.079 $\pm$  0.041 &  0.461 $\pm$  0.237\\
F470N  &  4.708 &  0.212 &  0.201 $\pm$  0.104 &  0.277 $\pm$  0.108 &  0.237 $\pm$  0.123 &  0.079 $\pm$  0.041 &  0.457 $\pm$  0.237\\
F480M  &  4.818 &  0.208 &  0.198 $\pm$  0.104 &  0.274 $\pm$  0.108 &  0.234 $\pm$  0.123 &  0.078 $\pm$  0.041 &  0.451 $\pm$  0.237\\
\hline
    \end{tabular}
    \tablefoot{Values based on the univariate spline interpolation obtained from the measured values in Table \ref{tab:results}. JWST NIRCam reference wavelengths using $\lambda_{\text{eff}} = \frac{\int F_\lambda P_\lambda \lambda^2 d\lambda}{\int F_\lambda P_\lambda \lambda d\lambda}$ for passband $P$ and using a $T_\text{eff} = 4750$ K star from \cite{CastelliKuruczModels}. The last two columns use $R_\text{F090W - F200W}$($V$) = 2.54 and $R_\text{F090W - F200W}$(${Ks}$) = 0.44, respectively.}
    \label{tab:NIRCam_values}
\end{table*}

\end{document}